\documentclass[12pt]{article}
\usepackage[dvips]{graphicx}
\usepackage{pdproc}

  \makeatletter 
  \def\@cite#1{[#1]} 
  \makeatother    
\begin{document}

\renewcommand{\thefootnote}{\alph{footnote}}

\title{
 Warped Supersymmetric Radius Stabilization\footnote{Talks 
 presented by N.M. at the 12th International Conference on Supersymmetry 
 and Unification of Fundamental Interactions, June 17-23, 2004, 
 Epochal Tsukuba, Tsukuba, Japan; 
 at YITP workshop 2004 "Progress in Particle Physics" 
 June 29-July 2, 2004, Yukawa Institute for Theoretical Physics 
 at Kyoto University, Japan; at Summer Insitute 2004, Aug 12-19, 2004, 
 Fuji-Yoshida, Japan.}
}

\author{Nobuhito Maru$^1$ and Nobuchika Okada$^2$}

\address{ 
$^1$ Theoretical Physics Laboratory, RIKEN \\
Wako, Saitama, 351-0198, JAPAN \\
$^2$ Theory Group, KEK \\
Tsukuba, Ibaraki, 305-0801, JAPAN
\\ {\rm E-mail: maru@riken.jp, okadan@post.kek.jp }}

\abstract{
A simple model of extra-dimensional radius stabilization in a supersymmetric 
Randall-Sundrum model is presented. 
In our model, we introduce only a bulk hypermultiplet 
and source terms on each boundary branes. 
With an appropriate choice of model parameters, 
we find that the radius can be stabilized by supersymmetric vacuum conditions. 
Since the radion mass can be much larger than the gravitino mass and 
the supersymmetry breaking scale, the radius stability is ensured even with 
the supersymmetry breaking. 
We find a parameter region where unwanted scalar masses induced by quantum 
corrections through the bulk hypermultiplet and a bulk gravity multiplet 
are suppressed and the anomaly mediation contributions dominate. 
}

\normalsize\baselineskip=15pt

\section{Introduction}
Recent attention to the brane world scenario is originally motivated 
by an alternative solution to the gauge hierarchy problem 
without supersymmetry (SUSY). 
There are two well-known scenarios, namely "Large Extra Dimension" \cite{ADD} 
and "Warped Extra Dimension" \cite{RS}. 
However, there is an alternative motivation to consider 
the brane world scenario in the context of SUSY breaking mediation 
in supergravity (SUGRA). 
In 4D SUSY phenomenology, SUSY particle spectrum is severely constrained 
to be almost flavor blind and CP invariant. 
In gravity mediation scenario, 
once SUSY is broken in the hidden sector, its breaking is transmitted to 
the visible sector through the Planck-suppressed SUGRA contact interaction 
between the hidden and the visible sector: 
\begin{equation}
\label{contact}
\int d^4 \theta c_{ij} \frac{Z^\dag Z Q^\dag_i Q_j}{M_4^2} 
\to c_{ij}m_{3/2}^2 \tilde{Q}_i^\dag \tilde{Q}_j, 
\end{equation}
where $Z$ is a hidden sector chiral superfield with nonvanishing 
vacuum expectation value (VEV) of auxiliary field, 
$Q_i$ is the minimal SUSY standard model (MSSM) chiral superfields of 
$i$-th flavor, $\tilde{Q}_i$ is its scalar component, 
$c_{ij}$ are flavor-dependent constants, and $M_4$ is a 4D Planck scale. 
If $F_Z \ne 0$, 
we obtain soft scalar masses of the order of the gravitino mass 
$m_{3/2}$ for the scalar partners. 
The problem is that there is no symmetry reason for $c_{ij} = \delta_{ij}$ 
in 4D SUGRA, namely $c_{ij} \ne \delta_{ij}$ in general. 
Therefore, the 4D SUGRA model suffers from 
the SUSY flavor problem.

Recently, it was proposed that the direct contact terms 
such as (\ref{contact}) are naturally suppressed if the visible sector and 
the hidden sector are separated from each other along the direction of extra 
spatial dimensions \cite{RS2,LS}. 
This is because the higher dimensional locality 
forbids the direct contact terms. 
In this setup, soft SUSY breaking terms in the visible sector are generated 
through a superconformal anomaly (anomaly mediation) and 
the resulting mass spectrum is found to be flavor blind \cite{RS2,GLMR}. 
Thus, it is well motivated to consider the SUSY brane world scenario.

In the brane world scenario, there is an important issue 
called "radius stabilization". 
For the scenario to be viable, the radius should be stabilized. 
In the normal brane world scenario, 
there is no radion potential in SUSY limit since the radion is a modulus. 
If SUSY is broken, the nontrivial potential is generated, 
such a potential usually destabilizes the radius. 
While we have to introduce additional bulk fields to stabilize the radius, 
these new fields might generate new flavor violating SUSY breaking terms 
larger than the anomaly mediation contributions \cite{LO}.  
This situation makes a realistic model construction very hard. 

\section{Model}
In this talk, we propose a simple model of extra-dimensional 
radius stabilization in a SUSY Randall-Sundrum model. 
Lagrangian of our model \cite{MO} is given by
\begin{eqnarray}
{\cal L}_5 &=& \int d^4 \theta \frac{T+T^\dagger}{2} 
e^{-(T+T^\dagger)\sigma}
\left[ -6 M_5^3 + |H|^2 + |H^c|^2 \right] |\phi|^2 \nonumber \\
&+& \left[ \int d^2 \theta \phi^3 e^{-3T \sigma} 
H \left\{
 \left( -\partial_y + \left( \frac{3}{2} + c \right) 
T \sigma' \right) H^c + W_b(y) \right\} + {\rm h.c.}
\right], \\
&\to&
\int d^4 \theta \left[ -3 M_5^3 (T+T^\dagger) |\omega|^2 
+ \frac{T+T^\dagger}{2}(|H|^2 + |H^c|^2) \right] \nonumber \\
&&+ \left[ \int d^2 \theta \omega H \left\{ - \partial_y H^c
+ \left( c + \frac{1}{2} \right) 
T \sigma' H^c + \omega W_b \right\} + {\rm h.c.} \right], 
\label{rescaled2}
\end{eqnarray}
where five dimensional spacetime metric compactified on $S^1/Z_2$ 
is given by 
\begin{equation}
ds^2 = e^{-2 r \sigma(y)} \eta_{\mu\nu}dx^\mu dx^\nu 
- r^2dy^2, ~(\mu, \nu=0,1,2,3)
\end{equation}
where $r$ is the radius of the fifth dimension, 
 $0 \leq y \leq \pi$ is the angle on $S^1$, 
 and  $\sigma(y) = k|y|$ with $k$ being an $AdS_5$ curvature scale. 
The prime denotes the differentiation 
with respect to $y$, $T$ is a radion chiral multiplet 
whose real part of scalar component gives the radius $r$, 
$\phi=1+\theta^2 F_\phi$ is a compensating multiplet, 
$H$ and $H^c$ are hypermultiplet components 
in terms of superfield notation in N=1 SUSY 
in four dimensions, 
and $Z_2$ parity for $H$ and $H^c$ are defined 
as even and odd, respectively. 
The boundary superpotentials are introduced as 
$W_b \equiv J_0 \delta(y) - J_\pi \delta(y-\pi)$, 
where $J_{0,\pi}$ are constant source terms 
on each boundary branes at $y = 0, \pi$. 
The second form of Lagrangian (\ref{rescaled2}) is obtained 
by making the field redefinition to simplify our analysis, 
\begin{equation}
(H, H^c) \to \frac{1}{\omega}(H, H^c), \quad \omega \equiv \phi e^{-T \sigma}. 
\end{equation}

It is straightforward to obtain SUSY solutions by solving F-flatness conditions,
\begin{eqnarray}
H(y) &=& C_H e^{(1/2-c)T \sigma} = 0,\quad 
H^c(y) \equiv \epsilon(y) \tilde{H}^c(y) = C_{\tilde{H}^c} 
\epsilon(y) e^{(c+1/2)T \sigma}, \\
\tilde{H}^c(0) &=& \frac{J_0}{2}, \quad \tilde{H}^c(\pi) = \frac{J_\pi}{2}
e^{-Tk \pi}, 
\end{eqnarray}
where $C_{H,\tilde{H}^c}$ are constants. 
These solutions give a SUSY vacuum condition of the form 
\begin{equation}
\label{SUSYvac}
0 = J_0 - J_\pi e^{-(3/2+c)Tk \pi}. 
\end{equation}
Thus, the radius is determined by appropriate values of $J_{0,\pi}$ 
and the bulk hypermultiplet mass $c$. 
This is our main result.

We have seen that the radius is stabilized, 
then we would like to know the radion mass. 
To do that, it is useful to describe our model 
in the form of 4D effective theory. 
Putting the SUSY classical solutions into our Lagrangian and integrating out 
the fifth dimensional coordinate, we can obtain 4D effective action. 
Now it is easy to calculate the radion potential and the radion mass.
\begin{eqnarray}
\label{potential}
V_{{\rm radion}} 
&=& \frac{(1-2c) k }{e^{(1/2-c)(T+T^\dag)k \pi}-1} 
\left| J_0 - J_\pi e^{-(c+\frac{3}{2}) Tk \pi} \right|^2, \\
\label{radionmass}
m^2_{{\rm radion}} &\sim& 
\frac{(1-2c)}{e^{(1/2-c)(T+T^\dag)k \pi}-1} 
\left(\frac{(\frac{3}{2}+c)^2 |J_\pi|^2}{3 M_5^3} \right) 
k^2 
\left. 
e^{-(1/2+c) (T+T^\dag) k \pi } \right|_{T=T_0} > 0. 
\end{eqnarray}
We can easily find the potential minimum 
$J_0 - J_\pi e^{-(c+\frac{3}{2}) T_0 k \pi} = 0$ from (\ref{potential}), 
which, of course, agrees with (\ref{SUSYvac}). 
Note that the radion mass squared is always positive, which means that 
the configuration we found is stable. 
As an example, 
if we take $c = 1/2$, $e^{-T_0 k \pi} \sim 10^{-2}, 
J_\pi \sim (0.1 \times M_5)^{3/2}$ and $k \sim 0.1 \times M_5$, 
we obtain the radion mass,  
\begin{equation}
\label{radionmass3}
m_{{\rm radion}}^2 \sim (10^{-5} \times M_4)^2 
 \gg m_{3/2}^2, F_{{\rm hidden}}, 
\end{equation}
which is much larger than the gravitino mass ($\sim 10$ TeV) 
in anomaly mediation scenario 
and the original SUSY breaking F-term scale 
$F_{{\rm hidden}}\sim m_{3/2} M_4 $ in a hidden sector. 
This fact implies that SUSY breaking effects little affect 
the radion potential, and the radius is not destabilized 
even in the presence of SUSY breaking. 
We have checked this expectation by considering SUSY breaking perturbatively. 
It is explicitly shown that the deviation of the radius from the radius in SUSY limit vanishes up to linear order of SUSY breaking scale.

In our model, we have a hypermultiplet and a gravity multiplet in the bulk. 
In general, there is a possibility that the hypermultiplet generates 
the flavor-dependent soft SUSY breaking sfermion masses 
since a zero mode of the parity even field $H$ can directly couple to 
both the hidden sector and the visible sector. 
The gravity multiplet can also couple to the both sectors and 
generates flavor-blind but tachyonic sfermion mass squareds \cite{GMSB}, 
which might break the standard model gauge group symmetry. 
It is in general possible that these contributions become larger than 
the anomaly mediation contributions. 
We can easily find the parameter region where these contributions to 
sfermion masses by the hypermultiplet and the gravity multiplet are 
strongly suppressed compared to the anomaly mediation spectrum. 
Therefore, we have no SUSY flavor problem. 

\section{Summary}
In this talk, we have proposed a simple model of extra-dimensional radius 
stabilization in the SUSY Randall-Sundrum model. 
With only a bulk hypermultiplet and the source terms on the boundary branes, 
the radius stabilization has been realized through SUSY vacuum conditions. 
The radion mass is found to be very large, 
this radius stabilization is ensured even 
if the SUSY breaking effects are taken into account.  
Our model gives a remarkable advantage 
for model building in the SUSY brane world, 
since the radius can be stabilized independently 
of the SUSY breaking and its mediation mechanism. 
Our model may be applicable to many models. 
We find a reasonable parameter region where 
unwanted contributions to scalar mass squared 
through the bulk multiplets are strongly suppressed.

\section{Acknowledgments}
We would like to thank the organizers of this conference for giving us
an opportunity to present this talk. 
N.M. is supported 
by Special Postdoctoral Researchers Program at RIKEN. 
N.O would like to thank the particle theory group 
of National Center for Theoretical Sciences (NCTS) in Taiwan 
for warm hospitality during his visit to NCTS 
where some parts of this work were completed.

\bibliographystyle{plain}

\begin{thebibliography}{99}
%
\bibitem{ADD}
 N.~Arkani-Hamed, S.~Dimopoulos and G.~R.~Dvali,
   Phys.\ Lett.\ B {\bf 429}, 263 (1998) [arXiv:hep-ph/9803315];
 I.~Antoniadis, N.~Arkani-Hamed, S.~Dimopoulos and G.~R.~Dvali,
   Phys.\ Lett.\ B {\bf 436}, 257 (1998) [arXiv:hep-ph/9804398]. 
%
\bibitem{RS}
%
 L.~Randall and R.~Sundrum,
  Phys.\ Rev.\ Lett.\  {\bf 83}, 3370 (1999) [arXiv:hep-ph/9905221];
  Phys.\ Rev.\ Lett.\  {\bf 83}, 4690 (1999) [arXiv:hep-th/9906064]. 
%
\bibitem{RS2}
 L.~Randall and R.~Sundrum,
  Nucl.\ Phys.\ B {\bf 557}, 79 (1999) [arXiv:hep-th/9810155]. 
%
\bibitem{LS}
 M.~A.~Luty and R.~Sundrum,
  Phys.\ Rev.\ D {\bf 62}, 035008 (2000) [arXiv:hep-th/9910202]. 
%
\bibitem{GLMR}
 G.~F.~Giudice, M.~A.~Luty, H.~Murayama and R.~Rattazzi,
  JHEP {\bf 9812}, 027 (1998) [arXiv:hep-ph/9810442]. 
%
\bibitem{LO}
 M.~A.~Luty and N.~Okada,
  JHEP {\bf 0304}, 050 (2003) [arXiv:hep-th/0209178]. 
%
\bibitem{MO} 
 N.~Maru and N.~Okada,
  Phys.\ Rev.\ D {\bf 70}, 025002 (2004) [arXiv:hep-th/0312148]. 
%
\bibitem{GMSB}
 T.~Gherghetta and A.~Riotto,
  Nucl.\ Phys.\ B {\bf 623}, 97 (2002) [arXiv:hep-th/0110022]; 
 I.~L.~Buchbinder, S.~J.~J.~Gates, H.~S.~J.~Goh, W.~D.~.~Linch, 
 M.~A.~Luty, S.~P.~Ng and J.~Phillips,
  Phys.\ Rev.\ D {\bf 70}, 025008 (2004) [arXiv:hep-th/0305169]; 
 R.~Rattazzi, C.~A.~Scrucca and A.~Strumia,
  Nucl.\ Phys.\ B {\bf 674}, 171 (2003) [arXiv:hep-th/0305184]. 
\end{thebibliography}

\end{document}